\documentstyle[epsf,preprint,tighten,aps]{revtex}
\begin{document}
\draft
\title{Temperature dependence of the nuclear symmetry energy}
\author{D. J. Dean, S. E. Koonin, K. Langanke, and P. B. Radha}
\address{W. K. Kellogg Radiation Laboratory, 106-38, California
Institute of Technology\\ Pasadena, California 91125 USA }
\date{\today}
\maketitle

\begin{abstract}
We have studied the properties of $A=54$ and $A=64$ isobars at
temperatures $T\le2$~MeV via Monte Carlo shell model calculations
with two different residual interactions. In accord with empirical
indications, we find that the symmetry energy coefficient, $b_{\rm
sym}$, is independent of temperature to within 0.6~MeV for
$T\le1$~MeV. This is in contrast to a recent suggestion of a 2.5~MeV
increase of $b_{\rm sym}$ for this temperature, which would have
significantly altered the supernova explosion scenario.
\end{abstract}
\pacs{}

\narrowtext
There appears to be a general consensus on the basic scenario for
presupernova collapse. An iron core is produced in the center of a
massive star in the final stage of hydrostatic nucleosynthesis. When
the iron core exceeds the Chandrasekhar mass limit, it can no longer
be supported by electron degeneracy pressure and so begins to
collapse. Subsequent photodisintegration of nuclei and electron
capture on nuclei (later on free protons) reduce the pressure and
energy of the core and accelerate its collapse, in this way raising
the central density and temperature. After neutrino trapping, the
inner part of the core (mass densities $\rho\gtrsim
10^{12}~\hbox{g/cm}^3$) collapses as a homologous unit whose size is
given by the appropriate Chandrasekhar mass limit, $M_{\rm Ch}
\approx 5.76 (Y_e^{\rm trap})^2 M_\odot$, where $Y_e^{\rm trap}$ is
the electron-to-nucleon ratio of the trapped material. The collapse
halts when the homologous core exceeds nuclear matter density
($\rho\gtrsim 10^{14}~\hbox{g/cm}^3$). A shock wave, formed at the
inner core's surface, then travels outward and eventually explodes
the star.

Despite the appeal of this scenario, the actual mechanism of a
supernova explosion is still controversial. In the {\it direct
mechanism} the shock wave is strong enough not only to stop the
collapse, but also to explode the outer shells. In the {\it delayed
mechanism} the shock energy is first stored in the core mantle (the
layer of infalling material in the iron core outside the homologous
region), but it is recovered shortly thereafter by the outstreaming
neutrinos. A detailed description of the physics of a supernova can
be found in Ref.~\cite{Bethe}.

The size of the core mantle is a major determinant of the supernova
mechanism. It has been argued recently that the mantle size might be
significantly smaller than generally calculated in supernova models
because of an overlooked temperature dependence of the electron
capture process \cite{Brown}. The detailed reasoning is as follows
\cite{Donati}: When the nucleus is described by a static mean field
model, dynamical effects associated with collective surface
vibrations are conventionally embodied in an effective nucleon mass,
$m^\ast$ (as are spatial non-localities in the mean field). Donati
{\it et al.} \cite{Donati} studied the coupling of the mean field
single-particle levels to the collective surface vibrations within
the quasiparticle random phase approximation and found that $m^\ast$
decreases with increasing temperature for $T\le2$~MeV, the
temperatures relevant for the presupernova collapse. They attributed
this behavior to a reduction of collectivity at low excitation
energies. In the Fermi gas model, this decrease of $m^\ast$ induces
an increase in the symmetry energy contribution to the nuclear
binding energy
\begin{equation}
E_{\rm sym} (T) = b_{\rm sym} (T) \frac{(N-Z)^2}{A}\;,
\end{equation}
where $N$, $Z$, and $A$ are the neutron, proton, and mass numbers of
the nucleus, respectively. Quantitatively, Donati {\it et al.}
estimate that $b_{\rm sym} (T)$ increases by about 2.5~MeV as $T$
increases from 0 to 1~MeV ($b_{\rm sym} (0)\approx28$~MeV
\cite{Bohr}).

Importantly, a larger value of $E_{\rm sym} (T)$ would hinder
electron capture in the presupernova environment, and thus increase
$Y_e^{\rm trap}$. This, in turn, would reduce the size of the core
mantle so that the shock wave would need significantly less energy to
stop the collapse and explode the star. In the supernova scenario
outlined by Donati {\it et al.}, the temperature dependence of
$b_{\rm sym}$ increases the shock energy by about 0.5~foe, roughly
half of the observed explosion energy of SN87a \cite{Donati}.

A more definitive test of the proposed temperature dependence of the
nuclear symmetry energy (and hence of the presupernova electron
capture process) necessarily involves a proper account of those
nuclear degrees of freedom beyond the mean field (e.g., pairing
correlations) that are relevant for nuclei under presupernova
conditions (e.g., $fp$-shell nuclei at $T\le2$~MeV). The model of
choice for this task is the interacting shell model, which provides a
complete nuclear spectrum and so allows a statistical description
using a canonical ensemble; both quantal and thermal fluctuations
(within the model space chosen) are fully included.

The recent development of the Monte Carlo shell model \cite{Lang} has
made feasible complete ($0\hbar \omega$) calculations of $fp$-shell
nuclei at finite temperature; a pilot study of the thermal properties
of the nucleus ${}^{54}$Fe has been presented in Ref.~\cite{Dean}. To
explore the temperature dependence of the symmetry energy, we have
used this method to study the thermal properties of two isobar chains
($A=54$ and 64) in the mass region important for the presupernova
collapse; this includes two of the three nuclei studied in
Ref.~\cite{Donati} (${}^{64}$Zn and ${}^{64}$Ni). Technical
considerations restrict our Monte Carlo shell model calculations to
the $N=Z$ and even-even members of the chains. The details of our
calculation parallel those of the ${}^{54}$Fe study in
Ref.~\cite{Dean} and the Monte Carlo shell model is described in
detail in Refs.~\cite{Lang,Alhassid}. For the $A=54$ isobar chain
(${}^{54}_{27}$Co, ${}^{54}_{28}$Fe, ${}^{54}_{30}$Cr, and
${}^{54}_{32}$Ti) we have adopted the Brown-Richter interaction
\cite{BR}. For the $A=64$ isobars (${}^{64}_{32}$Ge,
${}^{64}_{34}$Zn, ${}^{64}_{36}$Ni, and ${}^{64}_{38}$Fe), for which
the Brown-Richter force is expected to be unreliable \cite{Poves} (it
was constructed from properties of nuclei at the beginning of the
$fp$-shell), we have used the KB3 residual interaction \cite{Zuker}.
Further, to explore the sensitivity of our results to the two-body
interaction, we have also studied the $A=54$ chain with the KB3
interaction. All of our calculations omit the Coulomb interaction and
assume complete isospin symmetry.

\begin{minipage}[t]{6truein}{
$$\epsfxsize=5truein
\epsffile{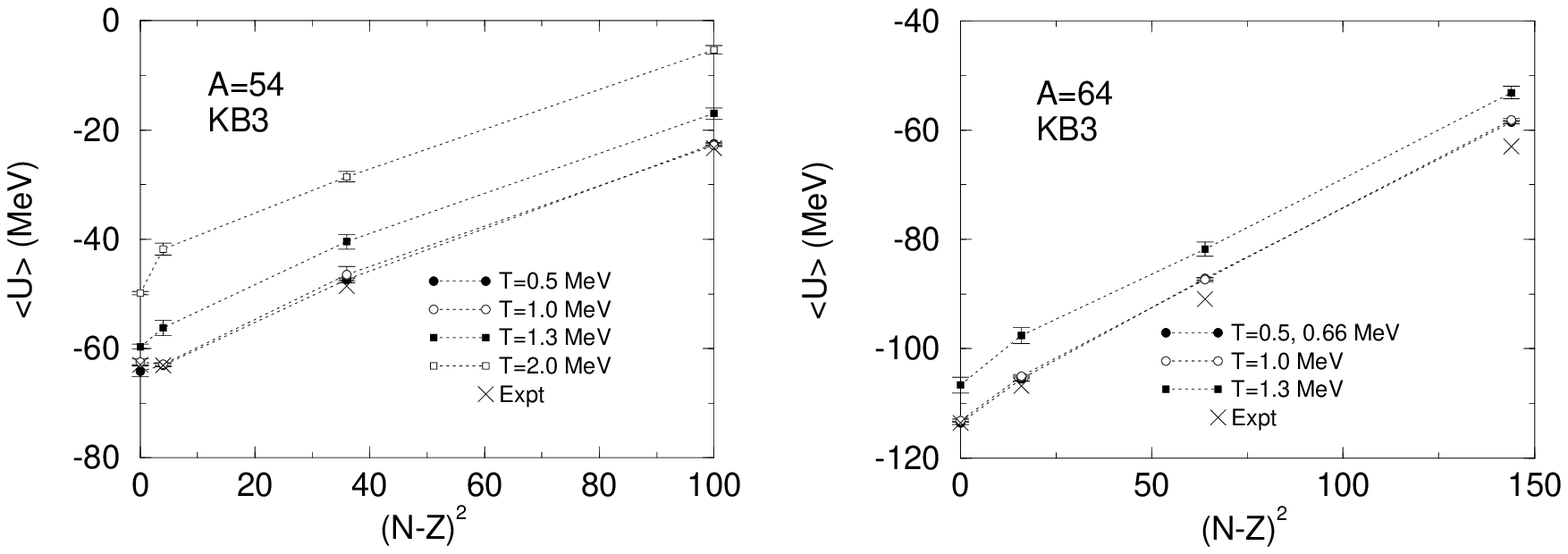}$$
\vbox{\footnotesize Figure 1. Calculated internal energies, $U$, for
the $A=54$ and $A=64$ isobar chains at various temperatures using the
KB3 interaction. Results for $A=54$ with the Brown-Richter
interaction are virtually identical. The energies are given relative
to the doubly-magic nucleus ${}^{40}$Ca. The experimental energies
have been corrected for the Coulomb repulsion as described in the
text and, in the case of $A=64$, have been increased by 11~MeV to
account for the systematic underbinding of the Hamiltonian for these
isobars.}}
\vspace{.5cm}
\end{minipage}

In Fig.~1 we display our results for $U$, the internal energy
(expectation value of the Hamiltonian in the canonical ensemble) for
the $A=54$ isobars and $A=64$ isobars as a function of $(N-Z)^2$ at
temperatures relevant for the presupernova. For even-even nuclei our
calculation at $T=0.5$~MeV has almost cooled to the ground state, as
the first excited states in these nuclei have excitation energies
$E_x\gtrsim 1$~MeV. For the $N=Z$ nucleus ${}^{54}$Co, the
$T=0.5$~MeV calculation represents a mixture of the ground state and
the first excited state at 200~keV. If we correct the experimental
binding energies by the semi-empirical Coulomb energy, $E_c= 0.717
Z^2/A^{1/3} (1-1.69/A^{2/3}$)~MeV \cite{Myers}, we find that our
calculation with either interaction accurately reproduces the $A=54$
binding energies. For the $A=64$ nuclei, the low-temperature
calculations are systematically overbound by some 11~MeV, although
the variation along the chain is well reproduced. A better
description of the nuclei ${}^{64}$Ni and ${}^{64}$Fe (with 16 and 18
valence neutrons in the $fp$-shell) would plausibly require inclusion
of the $g_{9/2}$ orbital. We observe that, at a given temperature,
$U$ is linearly proportional to $(N-Z)^2$ for the even-even nuclei of
an isobar chain. This behavior is, of course, expected from
semi-empirical parametrizations of the binding energy (e.g., the
Bethe-Weizs\"acker formula) and almost exclusively reflects the
symmetry energy contribution to the binding energies (1).

For $A=54$, our calculation yields $b_{\rm sym} (0)= 21\pm1.0$~MeV
and $22\pm0.75$~MeV for the KB3 and Brown-Richter interactions,
respectively. These values are in very good agreement with
experiment, $b_{\rm sym}= 21$~MeV. For $A=64$, we find $b_{\rm sym}=
23\pm0.7$~MeV, slightly larger than the experimental value of
$21.2$~MeV. We note that all of these values are significantly
smaller than the symmetry energy coefficient adopted in presupernova
studies ($b_{\rm sym}= 30$--32~MeV \cite{Brown,super}).

Our calculation does not confirm the temperature dependence of the
symmetry energy proposed in Ref.~\cite{Donati}. As is obvious from
Fig.~1, the slope of the linear variation of $U$ with $(N-Z)^2$ for
the even-even nuclei is independent of temperature, indicating that
$b_{\rm sym}$ does not change. To be more quantitative, an increase
of $b_{\rm sym}$ by 2.5~MeV between $T =0$ and 1~MeV, as proposed in
Ref.~\cite{Donati}, corresponds to an increase in the energy
splitting between ${}^{54}$Fe and ${}^{54}$Ti of $\Delta=4.4$~MeV
(using the experimental value for $b_{\rm sym}$), while our
calculation yields $\Delta= 0.4\pm0.9$~MeV, implying that $b_{\rm
sym}$ changes by less than 0.7~MeV. In fact, our calculation shows
that {\it all} properties of the even-even nuclei studied here are
essentially constant for $T\leq1$~MeV. These include the quadrupole
strength, whose variation with temperature has been cited as the
source for the proposed temperature-dependence of the symmetry energy
in Ref.~\cite{Donati}.

The low-temperature constancy of the properties of even-even nuclei
is expected in view of their sparse spectrum of low-lying excited
states and relatively high excitation energies of their first excited
states. To confirm this, we have calculated the nuclear
thermodynamics from the low-lying experimental levels of the nuclei
in the two isobar chains; enough of these levels are known to give
reliable results up to $T=0.7$~MeV. The partition function is given
by \begin{equation}
Z(T)=\sum_J (2J+1)e^{-E_J/T}\;,
\end{equation}
where the free energy is given by $F(T)= -T\ln Z(T)$, and the
internal energy is $U=-T^2\partial(F/T)/\partial T$. We then
calculated the difference in internal energies between isobars $a$
and $b$ as a function of the temperature, and expressed that in terms
of $b_{\rm sym}$ through Eq.~(1).

For the $A=54$ chain, we used all known levels in ${}^{54}$Fe and
${}^{54}$Cr up to 10~MeV of excitation energy; levels with unknown
spin were assigned a $J$-value randomly chosen from the standard
empirical spin distribution function \cite{Thielemann}. Results for
$U$ are shown in the left-hand panel of Fig.~2, expressed in terms of
$\Delta b_{\rm sym}$, the change in $b_{\rm sym}$ relative to its
value at $T=0$. There is structure in the curve due to the
peculiarities of the ${}^{54}$Fe and ${}^{54}$Cr spectra, although it
is smaller than the precision of our Monte Carlo calculations. We
find that $\vert b_{\rm sym}\vert$ changes by 0.3~MeV at most as $T$
varies from 0 to 0.7~MeV. A similar calculation in the $A=64$ chain
using ${}^{64}$Zn and ${}^{64}$Ni gives a change in $\vert b_{\rm
sym}\vert$ of 0.6~MeV as $T$ varies from 0 to 0.6~MeV. In contrast,
Donati {\it et al.} claim that $b_{\rm sym}$ increases by 1.22
(1.53)~MeV for ${}^{64}$Ni (${}^{64}$Zn) as $T$ varies from 0 to
0.7~MeV.\footnote{We have calculated the change of the symmetry
energy coefficient predicted by Donati {\em et al.} using the
exponential fit formula given in Table~I of Ref.~\cite{Donati}. We
note, however, that this formula significantly underestimates the
change in $b_{\rm sym}$ at $T=1$~MeV compared to the QRPA results
given in the same table.} Note that although we have defined the
symmetry energy coefficient in terms of $U$, rather than $F$, the
right-hand panel of Fig.~2 shows that the symmetry coefficient of $F$
never varies by more than 0.5~MeV for $T<0.7$~MeV. It is also
interesting to note that, depending on the low-lying spectra, $b_{\rm
sym}$ may either increase or decrease as $T$ increases.

\begin{minipage}[t]{6truein}{
$$\epsfxsize=5truein
\epsffile{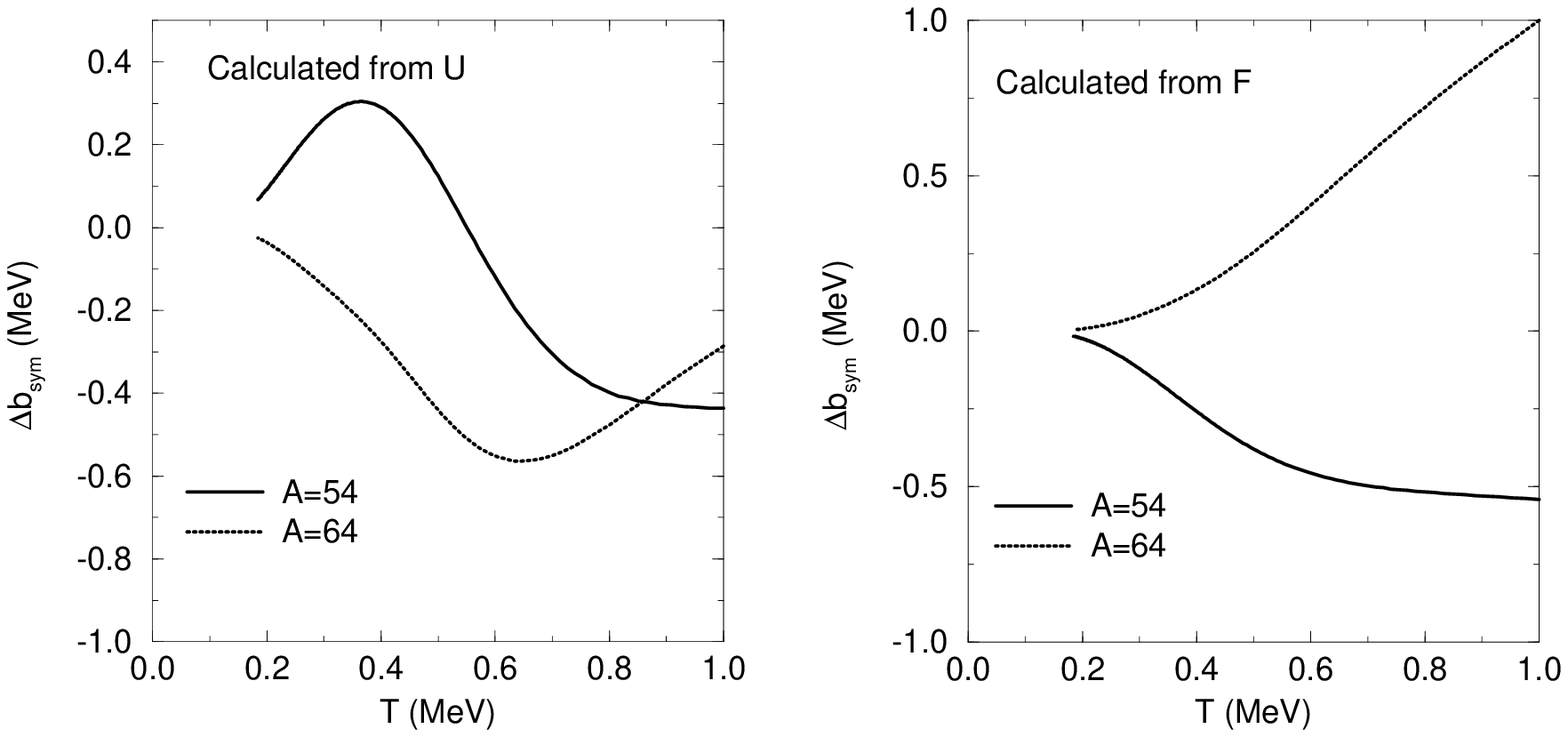}$$
\vbox{\footnotesize Figure 2. Temperature dependence of the symmetry
energy coefficients for the internal energy, $U$, and free energy,
$F$, calculated from the empirical spectra as described in the text.
${}^{54}$Fe and ${}^{54}$Cr were compared to calculate the $A=54$
case, while ${}^{64}$Zn and ${}^{64}$Ni were used for $A=64$. The
limited knowledge of levels at high excitation energy makes these
cases unreliable for $T\protect\gtrsim 0.7$~MeV.}}
\vspace{.5cm}
\end{minipage}

In contrast to even-even nuclei, odd-odd nuclei (with $N$ not equal
to $Z$) have a rather rich spectrum at low excitation energies. Thus,
$U$ is expected to increase more strongly for these nuclei at
temperatures $T<1$~MeV. From a simple Fermi gas picture one expects
an increase of $a T^2\approx 7$~MeV, with the empirical level density
parameter $a=A/8~\hbox{MeV}^{-1}$ \cite{Thielemann}. The increase of
$U$ with temperature for odd-odd nuclei changes the kinematics of
electron capture, making it easier for odd-odd nuclei to capture
electrons.

As is demonstrated in detail in Ref.~\cite{Dean} for the case of
${}^{54}$Fe, our Monte Carlo shell model calculations predict the
disappearance of pairing correlations between like nucleons at
$T\approx 1.3$~MeV, while $np$-correlations persist to higher
temperatures. We observe the same behavior for the nuclei studied
here. At $T\approx 1.3$~MeV we find a significant change in the
thermal properties of the nuclei. For example, the moment of inertia
and the (orbital) $M1$ strength increase drastically. This change is
also apparent from the increase of $U$ with temperature, as shown in
Fig.~1. Monitored by the number of BCS-like pairs in the nuclei, we
observe that the change is related to the disappearance of the $pp$-
and $nn$-correlations, which, in turn, dominate the nuclear
properties at low temperatures. The vanishing of the pairing energy
in even-even nuclei at $T\geq 1.3$~MeV has a significant effect on
the kinematics of the electron capture on nuclei in the late stage of
presupernova collapse.

Another novel feature is apparent in our calculations, although it
has no direct relation to the presupernova situation. It is well
known that odd-odd $N=Z$ nuclei are slightly less bound relative to
the even-even nuclei in an isobar chain because of the lack of
like-particle pairing of the last proton and neutron. However, there
is apparently a strong correlation between these unpaired nucleons in
$N=Z$ nuclei that energetically favors them over the other odd-odd
nuclei in the chain. For both interactions, we observe in the $A=54$
isobar chain that ${}^{54}$Co becomes energetically favored relative
to the even-even nuclei, after the disappearance of the like-pair
correlations in the latter. This observation is in agreement with the
finding in \cite{Dean} that isoscalar $np$-correlations persist to
higher temperatures than the like-pair correlations. Apparently these
correlations still contribute to the internal energy of ${}^{54}$Co
at $T\geq 1.3$~MeV.

In conclusion, Monte Carlo shell model calculations show clearly that
pairing correlations play a decisive role in nuclear properties at
low temperatures. For all ($N=Z$ and even-even) $fp$-shell nuclei
studied, we find that the nuclear properties are essentially constant
for $T\leq1$~MeV, as expected from the spectrum of these nuclei. In
particular, we find that the symmetry energy coefficient is constant
to within some 0.6~MeV for $T<1$~MeV. Thus, our calculation does not
confirm speculations based on a mean-field approximation that the
symmetry energy increases by 2.5~MeV in the temperature interval
$T=0$--1~MeV. Our calculations illustrate that nucleon-nucleon
correlations beyond the mean-field level must be taken into account.
A detailed account of this work will be given elsewhere.

This work was supported in part by the National Science Foundation,
Grants No. PHY90-13248 and PHY91-15574. We thank the Concurrent
Supercomputing Consortium for a grant of DELTA and PARAGON time.

\end{document}